\documentclass[12pt,a4paper,final]{iopart}

\usepackage{iopams}  
\usepackage{graphicx}
\usepackage[breaklinks=true,colorlinks=true,linkcolor=blue,urlcolor=blue,citecolor=blue]{hyperref}
\usepackage{braket}
\usepackage{caption}
\usepackage{abstract}
\usepackage{float}

\expandafter\let\csname equation*\endcsname\relax

\expandafter\let\csname endequation*\endcsname\relax

\usepackage{amsmath}

\begin{document}

\title[P.W.K. Jensen et al./Molecular Realization of a Quantum \texttt{NAND} Tree]{Molecular Realization of a Quantum \texttt{NAND} Tree}

\author{Phillip W.K. Jensen$^{1,2}$}
\address{$^1$Nano-Science Center and Department of Chemistry, University of Copenhagen, Universitetsparken 5, 2100 Copenhagen, Denmark.}
\address{$^2$Department of Chemistry and Chemical Biology, Harvard University, 12 Oxford St., Cambridge, MA 02138, United States.}

\ead{phillip.kastberg@gmail.com}

\author{Chengjun Jin$^{1}$}
\address{$^1$Nano-Science Center and Department of Chemistry, University of Copenhagen, Universitetsparken 5, 2100 Copenhagen, Denmark.}
\ead{jin.chengjun@icloud.com}

\author{Pierre-Luc Dallaire-Demers$^{2,3}$}
\address{$^2$Department of Chemistry and Chemical Biology, Harvard University, 12 Oxford St., Cambridge, MA 02138, United States.}
\address{$^3$Xanadu, 372 Richmond St W, Toronto, M5V 2L7, Canada.}
\ead{dallairedemers@gmail.com}

\author{Al\'{a}n Aspuru-Guzik$^{4,5,6}$}
\address{$^4$Department of Chemistry and Department of Computer Science, University of Toronto, 80 St. George Street, Toronto, Ontario M5S 3H6, Canada.}
\address{$^5$Canadian Institute for Advanced Research (CIFAR) Senior Fellow, 661 University Ave., Suite 505, Toronto, ON M5G 1M1, Canada.}
\address{$^6$Vector Institute, 661 University Ave., Suite 710 Toronto, ON M5G 1M1, Canada.}
\ead{alan@aspuru.com}

\author{Gemma C. Solomon$^{1}$}
\address{$^1$Nano-Science Center and Department of Chemistry, University of Copenhagen, Universitetsparken 5, 2100 Copenhagen, Denmark.}
\ead{gsolomon@chem.ku.dk}

\begin{abstract}
The negative-\texttt{AND} (\texttt{NAND}) gate is universal for classical computation making it an important target for development. A seminal quantum computing algorithm by Farhi, Goldstone and Gutmann has demonstrated its realization by means of quantum scattering yielding a quantum algorithm that evaluates the output faster than any classical algorithm. Here, we derive the \texttt{NAND} outputs analytically from scattering theory using a tight-binding (TB) model and show the restrictions on the TB parameters in order to still maintain the \texttt{NAND} gate function. We map the quantum \texttt{NAND} tree onto a conjugated molecular system, and compare the \texttt{NAND} output with non-equilibrium Green's function (NEGF) transport calculations using density functional theory (DFT) and TB Hamiltonians for the electronic structure. Further, we extend our molecular platform to show other classical gates that can be realized for quantum computing by scattering on graphs. 
\end{abstract}

\vspace{2pc}

\section{Introduction}

\begin{figure}[H] 
\centering  
\includegraphics[width=0.7\textwidth]{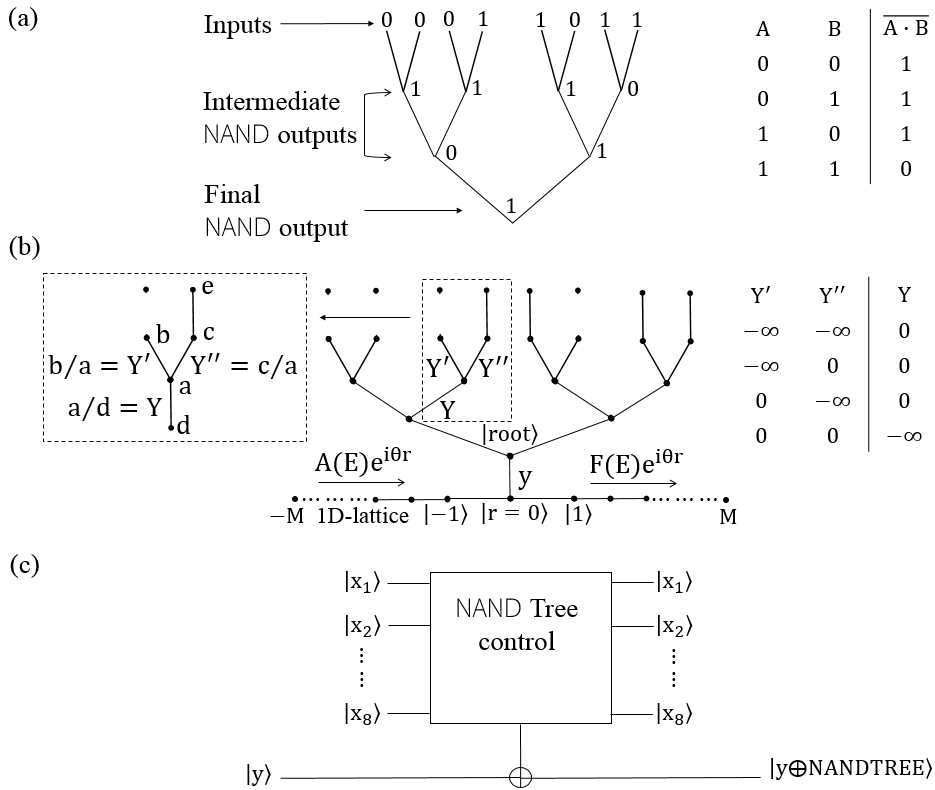} 
\caption{ (a) A classical \texttt{NAND} tree with four inputs,  intermediate \texttt{NAND} outputs and a final \texttt{NAND} output. The truth table for a classical \texttt{NAND} gate (right), where the \texttt{NAND} output, $\overline{\text{A} \cdot \text{B}}$, is the product of A and B followed by the logical complement (flipping the bit). (b) A quantum \texttt{NAND} tree where each node (black dot) correspond to a site in the tight-binding Hamiltonian and the 1D-lattice is running from -M to M where we will take M to be very large. A particle is propagating from left to right, where the amplitudes of the incident, reflected, and transmitted particle are given as $A(E) e^{i \theta r}$, $B(E) e^{-i \theta r}$, and $F(E) e^{i \theta r'}$, respectively. The quantum \texttt{NAND} tree, illustrated here, has four inputs $\{ Y' , Y'' \}$, controlled by connection/disconnection states at the top of the tree, intermediate \texttt{NAND} outputs $\{Y\}$, and a final \texttt{NAND} output $\{y\}$. The inputs and \texttt{NAND} outputs are given as amplitude ratios between two connection states, input($Y'=b/a, Y''=c/a$) and \texttt{NAND} output ($Y=a/d$), where the letters are amplitudes, $a := \braket{a|E}, \hdots , e := \braket{e|E}$, at a specific node. The truth table for a quantum \texttt{NAND} gate (right) with bits $\{-\infty, 0\}$, and setting $-\infty := 0$ and $0 := 1$, renders the truth table for the quantum \texttt{NAND} gate equivalent to its classical counterpart. (c) Quantum circuit representation of the \texttt{NAND} tree.}
 \label{fig:introduction}
\end{figure}

Quantum computers process information according to the principles of quantum mechanics with the advantage of solving certain problems, such as binary addition and factoring integers\cite{Deutsch1992}\cite{Shor1995}, faster than classical computers. Quantum random walks have been suggested as a key component in the development of quantum algorithms for quantum computing\cite{Farhi1998}\cite{Farhi2007}\cite{ChildsPRL2009}\cite{Childs2009a}\cite{Childs2011}\cite{Childs2012}\cite{Whitfield2010}. Any evolution in a finite-dimensional Hilbert space can be thought of as a quantum random walk, however,  it is clearest when the Hamiltonian has a local structure (see Eq. (\ref{Eq. 1})). An example of this approach is the seminal work of Farhi et al.\cite{Farhi2007} which shows how the classical \texttt{NAND} tree, illustrated in Fig. \ref{fig:introduction}(a), can be realized for quantum computing by scattering on graphs. A \texttt{NAND} tree, a composition of \texttt{NAND} gates, produces a one bit output at the end of the computation and any Boolean function, $f: \{0,1\}^n \rightarrow \{0,1\}$, that has output $m$ bits can be computed from $m$ \texttt{NAND} trees alone.  Hence, any binary calculation (addition, subtraction and multiplication) can be computed using \texttt{NAND} trees.

\hfill \break
The classical \texttt{NAND} tree has inputs, intermediate \texttt{NAND} outputs and a final \texttt{NAND} output, as shown in Fig. \ref{fig:introduction}(a). The inputs come in pairs and for the classical \texttt{NAND} tree, illustrated here, the inputs are $(0,0), (0,1), (1,0), (1,1)$ given at the leaves. In general, the number of inputs is $2^n/2$, where $n$ is the depth of the \texttt{NAND} tree (Fig. \ref{fig:introduction}(a) has depth 3). The intermediate \texttt{NAND} outputs are given between the inputs and final \texttt{NAND} output, thus in the example of Fig. \ref{fig:introduction}(a) the intermediate \texttt{NAND} outputs are (1,1), (1,0), and (0,1), and the final \texttt{NAND} output, the bit at the end of the computation, is (1). Farhi et al.'s\cite{Farhi2007} schematic representation of the quantum \texttt{NAND} tree is illustrated in Fig. \ref{fig:introduction}(b), where each node corresponds to a site in the tight-binding (TB) Hamiltonian. As in the classical case, the inputs to the quantum \texttt{NAND} tree $\{Y',Y''\}$ are given at the leaves and we control the inputs by connecting/disconnecting states. The intermediate \texttt{NAND} outputs $\{Y\}$ are given between the inputs and the final \texttt{NAND} output, and the final \texttt{NAND} output $\{y\}$ is given at the bottom. The bits in the quantum truth table are $\{-\infty, 0\}$, and setting $-\infty := 0$ and $0 := 1$, renders the truth table for the quantum \texttt{NAND} gate equivalent to its classical counterpart. For the quantum \texttt{NAND} tree, we measure the final \texttt{NAND} output, $\{y\}$, in the transmission coefficient. If the incoming particle is fully reflected the final \texttt{NAND} output is (0), or if the particle is fully transmitted the final \texttt{NAND} output is (1). Farhi et al. show that the run time to evaluate the quantum \texttt{NAND} tree, with a quantum algorithm, has a lower bound of $\sqrt[]{N}$ with $N=2^n$ (number of leaves), where the run time for the best known classical algorithm is $N^{0.753}$\cite{Farhi2007}. Thus, the quantum algorithm evaluates the final \texttt{NAND} output faster, which could enable investigation of systems, that are impossible, or limited, with classical algorithms. Fig. \ref{fig:introduction}(c) shows the reversible circuit representation of the quantum \texttt{NAND} tree where the qubits $\ket{x_1},\ket{x_2},\hdots,\ket{x_8}$ work as inputs to the quantum \texttt{NAND} tree. The quantum \texttt{NAND} tree operates as a sequence of multi-control \texttt{NOT} gates where the bit flip acts on the output register.
\hfill \break

\section{Theory}
For the quantum \texttt{NAND} tree, as suggested by Farhi et al.\cite{Farhi2007}, we describe the system with a TB Hamiltonian with nearest-neighbor interactions. The single-particle Hamiltonian in the site representation, $\{\ket{r}\}$, of the atomic orbitals reads,

\begin{equation}
\hat{\text{H}}= \sum_{\text{r=-M}}^{\text{M}}   \alpha \ket{r}\bra{r} + \sum_{\text{r=-M}}^{\text{M}}  \beta \big(\ket{r}\bra{r + 1}+ \ket{r + 1}\bra{r}\big) +\hat{\text{H}}_{\text{TREE}}.
\label{Eq. 1}
\end{equation}
The first three terms sum over the states attached to the semi-infinite 1D-lattice with site energies and coupling elements given as $\alpha$ and $\beta$, respectively. The discrete basis states $\ket{r}$, where $r = \text{-M},\hdots,\text{M}$ are the lattice states and we will take M to be very large. The last term, $\hat{\text{H}}_{\text{TREE}}$, is the TB Hamiltonian for the \texttt{NAND} tree and is defined similarly to the first three terms with only nearest-neighbour interactions included. For a semi-infinite 1D-lattice, the energy eigenstate and eigenvalue, $\hat{\text{H}}_ {\text{1D}}\ket{E} = E \ket{E}$, are given by $\ket{E}=A(E)\sum_{r} e^{i\theta r} \ket{r} + B(E)\sum_{r} e^{-i\theta r} \ket{r}$ and $E=\alpha + 2\beta \cos(\theta)$, respectively, with $\theta$ as the incoming momentum. For each $\theta \in \mathbb{R} $, there is a scattering eigenstate $\ket{E}$ of momentum $\theta$, and from standard quantum scattering theory (QST), the scattering eigenstates are given by,

\begin{align}
\braket{r|E} &= \big[A(E)e^{ir\theta}+ B(E)e^{-ir\theta} \big]  \quad \text{for } r \leq 0,\label{Eq.2}\\
\braket{r|E}&= F(E)e^{ir\theta} \hspace{3.1cm} \text{for} \hspace{0.1cm}  r \geq 0, \label{Eq.3}
\end{align}
where the \texttt{NAND} tree is located at $r=0$ and $A(E)e^{ir\theta}$, $B(E)e^{-ir\theta}$, and $ F(E)e^{ir'\theta}$ are amplitudes of the incident, reflected, and transmitted particle, respectively, expressed in the $\{\ket{r}\}$-basis (site representation). The amplitude function of the transmitted particle is given by,

\begin{equation}
F(E)= A(E) \frac{2i \sin(\theta)}{2i\sin(\theta) + y(E)}, \quad y(E) = \frac{\braket{\mbox{root}|E}}{\braket{r=0|E}},
\label{Eq.4}
\end{equation}
where the states $\ket{r=0}$ and $\ket{\text{root}}$ are illustrated in Fig. \ref{fig:introduction}(b). We have assumed equal site energies and coupling elements, setting $\alpha = 0$ eV and $\beta=-1.0$ eV (see supporting information S1). The quantity $y(E)$ is an amplitude ratio of the scattering eigenstate $\ket{E}$ between the states $\ket{\mbox{root}}$ and $\ket{r=0}$ and gives the final \texttt{NAND} output for the quantum \texttt{NAND} tree. All information regarding the tree i.e., the structure of the tree, coupling elements, site energies etc., are contained in $y(E)$. The transmission function is calculated from $T(E)=|F(E)|^2/|A(E)|^2$. Letting $E \rightarrow 0$, depending on the leaf structure (inputs), the final \texttt{NAND} output, $y(E)_{E \rightarrow 0}$, takes the binary values,

\begin{equation}
y(E)_{E \rightarrow 0 \hspace{0.1cm}} =
\begin{cases} 
    - \infty,      & \quad F(E)_{E \rightarrow 0} = 0 \quad \quad  \hspace{1.00cm} \Rightarrow \quad  T(E)_{E \rightarrow 0} = 0 \hspace{0.40cm} \texttt{NAND} \hspace{0.10cm} (0).  \\
    \hspace{0.30cm}0,  & \quad F(E)_{E \rightarrow 0} = A(E)_{E \rightarrow 0} \quad \Rightarrow \quad T(E)_{E \rightarrow 0} = 1 \hspace{0.41cm} \texttt{NAND} \hspace{0.10cm} (1).
   \end{cases}
\label{Eq.1}
\end{equation}
For the final \texttt{NAND} output $y(E)_{E \rightarrow 0} = -\infty$, the transmitted amplitude coefficient equals to zero, $ F(E)_{E \rightarrow 0} = 0$, and we measure the final \texttt{NAND} output (0) with a fully reflected particle. For the final \texttt{NAND} output $y(E)_{E \rightarrow 0} =0$, the transmitted amplitude coefficient equals the incident, $F(E)_{E \rightarrow 0} = A(E)_{E \rightarrow 0}$, and we measure the final \texttt{NAND} output (1) with a fully transmitted particle. We evaluate the quantum \texttt{NAND} tree by starting from the inputs $\{Y',Y''\}$ and moving down the tree, as in the classical \texttt{NAND} tree, and the bit values, $\{-\infty, 0\}$, are the result of amplitude ratios between two connection states and by letting $E \rightarrow 0$. For example, the input and \texttt{NAND} output for the leaf structure highlighted in Fig. \ref{fig:introduction}(b) are given by (see supporting information S2),
\begin{align}
&\underbrace{Y'(E)=\frac{b}{a} = -\frac{1}{E}, \quad Y''(E)=\frac{c}{a} = \frac{E}{1-E^2}}_{\text{input}(Y',Y'')}, \quad \underbrace{Y(E)= \frac{a}{d} = \frac{-1}{E + Y'(E)+Y''(E)}}_{\texttt{NAND} \hspace{0.1cm} \text{output}}.
\label{Eq.InpOut}
\end{align}
Letting $E \rightarrow 0^+$, the input becomes $Y'=-\infty$ and $Y''=0$, which results in the \texttt{NAND} output ($Y=0$). The other inputs, $\{-\infty, -\infty \}$, $\{0, -\infty \}$, and $\{0, 0 \}$, differ in leaf structure.
\hfill \break 
\hfill \break 
While organic molecules are not normally included in systems proposed for quantum circuits, when the quantum random walk (the evolution of our system, Eq. (\ref{Eq. 1}), can be thought of as a quantum random walk) is recast as a scattering problem the formulation is reminiscent of systems previously studied in molecular electronics. When a conjugated organic molecule is modelled with a tight-binding Hamiltonian, each carbon atom can be represented by a single site and the nearest-neighbour interactions are determined by the nature of the carbon-carbon bonds. Imagining the tight-binding system as a molecule puts some restrictions on the topology and the assumptions outlined above need to be lifted to approach a chemically sensible system. In most cases, a molecule cannot be treated as having the same coupling elements and equal site energies across its entirely. Here, our goal is to realize the quantum \texttt{NAND} tree in a molecular system given realistic physical limitations of molecular platforms. Earlier work  realizing Boolean truth tables with molecules using the Quantum Hamiltonian computation approach\cite{Dridi2015}\cite{Dridi2018} has employed organic molecules and TB Hamiltonians for molecules on a surface\cite{Kleshchonok2015}\cite{Joachim2012}\cite{Soe2011}\cite{Joachim2005}.

\section{Results and discussion}
\begin{figure}[h] 
\centering 
\includegraphics[width=1.0\textwidth]{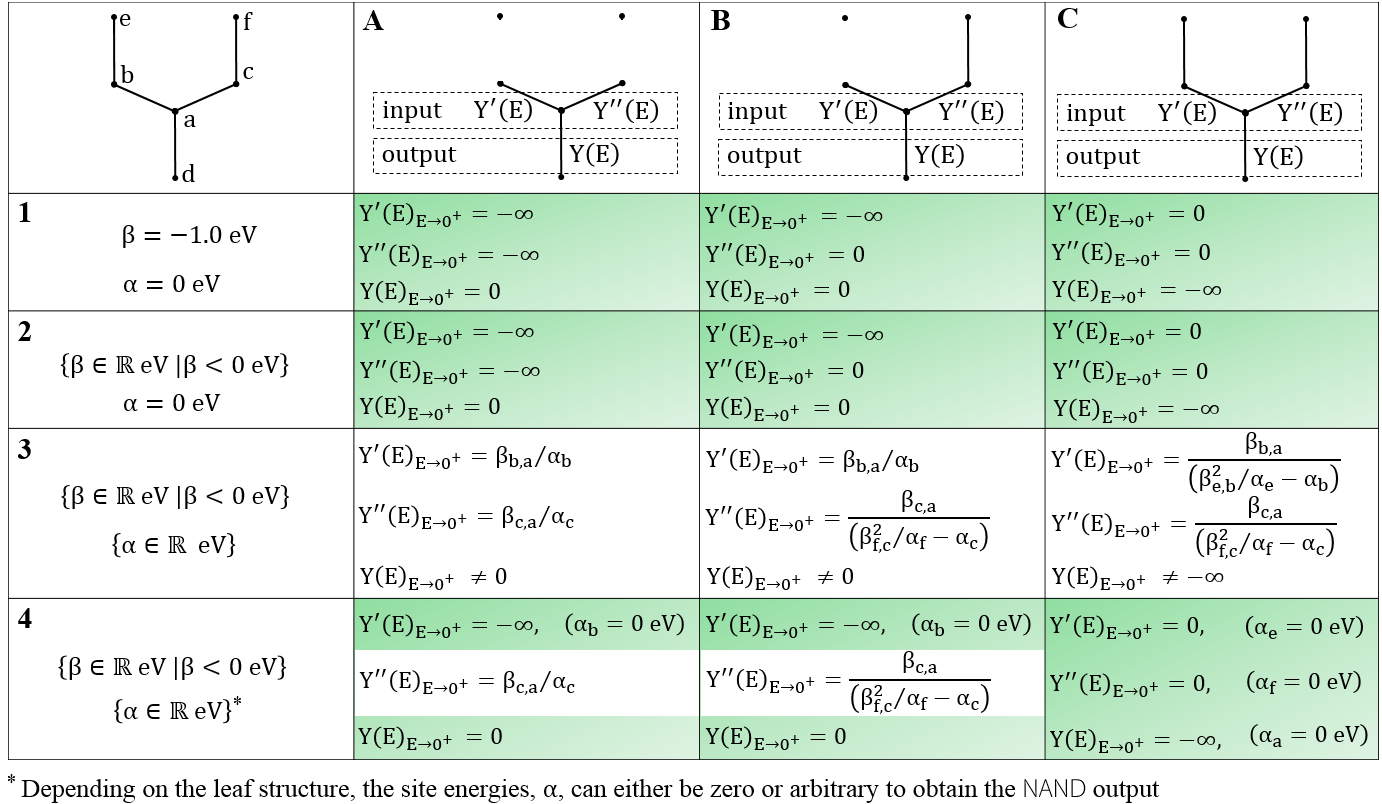} 
\caption{Inputs and \texttt{NAND} outputs derived from standard quantum scattering theory. Row 1, column 1: The letters located at each node, $a, \hdots,f$, are amplitudes of the eigenstate $\ket{E}$ at the corresponding node that is $a \equiv \braket{a|E}, \hdots, f \equiv \braket{f|E}$. Cases A,B,C: Three different leaf structures corresponding to different inputs to the quantum \texttt{NAND} tree. The input($Y',Y''$) is given as amplitude ratios between two connection states, $ Y'(E)=b/a$ and $ Y''(E)=c/a$, and the \texttt{NAND} output is given as the amplitude ratio $Y(E)=a/d$. If node $b$ is only connected to node $a$ then $Y'(E)=b/a=\beta_{b,a}/ \big(E-\alpha_ {b}\big)$, or if node $b$ is connected to node $a$ and $e$ then $Y'(E)=b/a =  E \beta_{b,a} / \big( \big(E^2-E \beta_{e,b}^2 \big) / \big(E- \alpha_ {e}\big) -E \alpha_ {b} \big)$. The same expressions are used for $Y''(E)=c/a$. The \texttt{NAND} output is given as $Y(E)=a/d =\beta_{a,d} / \big(E-\alpha_{a} - \beta_{b,a}Y'(E)-\beta_{c,a}Y''(E)\big)$. Cases 1-4: Differ in their restrictions on coupling elements, $\beta$, and site energies, $\alpha$. Shaded green backgrounds indicate that bit values $\{-\infty,0\}$ are preserved which are required in order to have a \texttt{NAND} gate function preserved in $Y(E)$.} 
\label{table:theory}
\end{figure}

The first question as we move from the model to a molecule is whether we still have the value of a \texttt{NAND} gate function encoded in the transmission coefficient if we include variation in coupling elements and site energies throughout a quantum \texttt{NAND} tree. In Fig. \ref{table:theory}, we show three different leaf structures, A, B, C, that are three different inputs to the quantum \texttt{NAND} gate and their corresponding \texttt{NAND} outputs. As previously mentioned, the input($Y',Y''$) is given as the amplitude ratio between two connection states, $ Y'(E)=b/a$ and $ Y''(E)=c/a$, and the \texttt{NAND} output is given as the amplitude ratio $Y(E)=a/d$. Cases 1-4 differ in their restrictions on coupling elements and site energies, and shaded green backgrounds indicate that bit values $\{-\infty,0\}$ are preserved. The bit values are required in order to have a \texttt{NAND} gate function preserved in $Y(E)$. In the simplest situation, case 1, coupling elements and site energies are kept constant with $\beta=-1.0$ eV and $\alpha = 0$ eV, which is the assumption we have used up to this point. For $E \rightarrow 0^+$, all inputs are bit values, hence the \texttt{NAND} gate function is preserved in $Y(E)$. In case 2, we consider arbitrary negative coupling elements but site energies are kept constant. For $E \rightarrow 0^+$, all inputs are bit values, hence the \texttt{NAND} gate function is preserved in $Y(E)$ and is independent of coupling strengths (see supporting information S2). 

\hfill \break
Next we evaluate the effect of variation in site energies. In case 3, we consider arbitrary negative coupling elements and arbitrary site energies, thus we have removed all restrictions. For $E \rightarrow 0^+$, none of the bit values are preserved and the \texttt{NAND} gate function is not preserved in $Y(E)$ (see supporting information S3). In case 4, we tighten the restrictions again, considering arbitrary negative coupling elements but, depending on the leaf structure, the site energies are either restricted to zero or allowed to take an arbitrary value. For example, the leaf structure A should, in the ideal case, have input$(Y'=Y''=-\infty)$, which results in \texttt{NAND} output $(Y=0)$. However, this is only possible if all site energies, $\alpha_b$ and $\alpha_c$, are zero. If we change only one of the site energies, for example $\alpha_c \neq 0$ eV, then the input$(Y'=-\infty,Y''=\mathbb{R})$ still results in \texttt{NAND} output $(Y=0)$, since the input($Y',Y''$) is located in the denominator in the \texttt{NAND} output, $Y(E) =\beta_{a,d} / \big(E-\alpha_{a} - \beta_{b,a}Y'(E)-\beta_{c,a}Y''(E)\big)$. Hence, for the leaf structure A, we can change one of the two site energies, $\alpha_b$ or $\alpha_c$, and still preserve the \texttt{NAND} output ($Y=0$). For the leaf structure B, the site energy $\alpha_b$ is restricted to zero to preserve the \texttt{NAND} output ($Y =0$). For the leaf structure C that has the \texttt{NAND} output ($Y=-\infty$), the three site energies $\alpha_e$, $\alpha_f$, and $\alpha_a$ are all restricted to zero to preserve the \texttt{NAND} output ($Y = -\infty$).
\hfill \break
\hfill \break
From these analytical results, we can conclude that the \texttt{NAND} gate function can be maintained without restrictions on coupling elements, and site energies can be varied, depending on the input. With this relaxation of assumptions, we will build a model molecular system for the quantum \texttt{NAND} tree. We can map the TB model onto a conjugated molecular system, where we treat each site as a carbon or nitrogen atom in the $\pi$-system and vary the coupling elements to model the single and double bonds in the molecule.

\begin{figure}[h] 
\centering  
\includegraphics[width=0.75\textwidth]{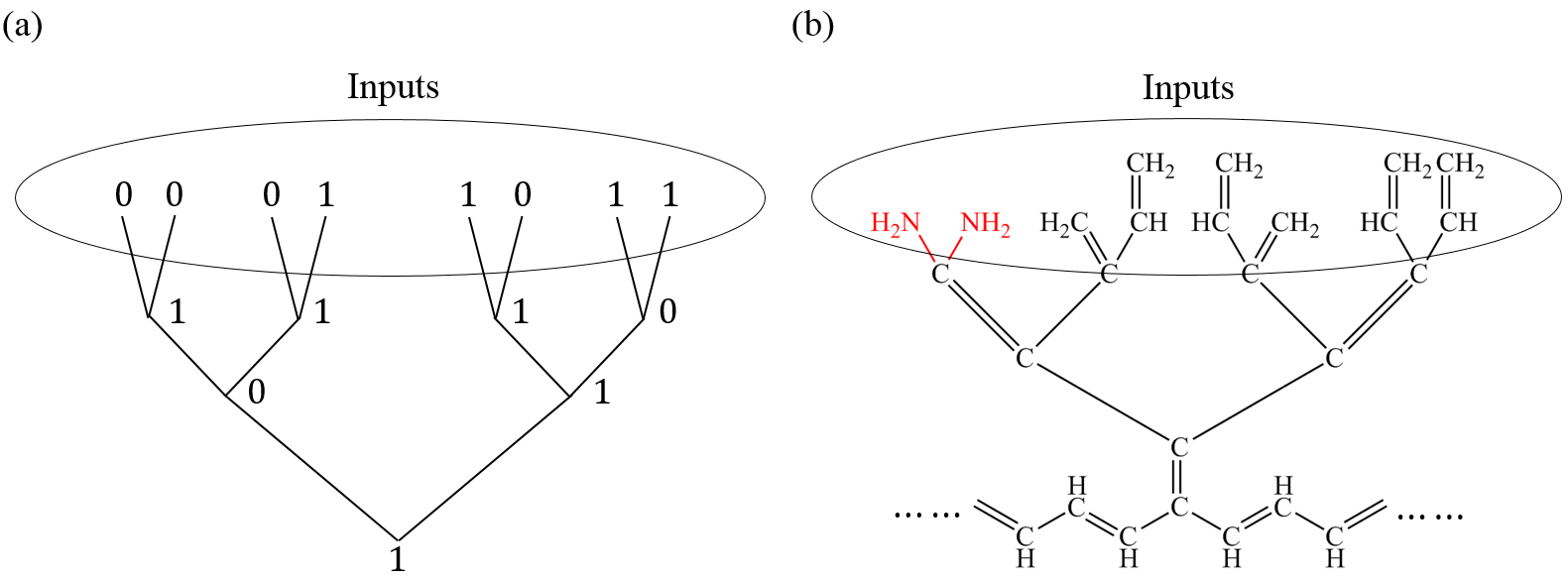} 
\caption{A molecular system for the quantum \texttt{NAND} tree. (a) The classical \texttt{NAND} tree. (b) A molecular system for the quantum \texttt{NAND} tree, analogous to the classical \texttt{NAND} tree in (a) by having the same inputs. We have added two nitrogen atoms at the leaves to avoid unstable carbon radicals.}
\label{model1}
\end{figure}
\begin{figure}[h] 
\centering  
\includegraphics[width=0.9\textwidth]{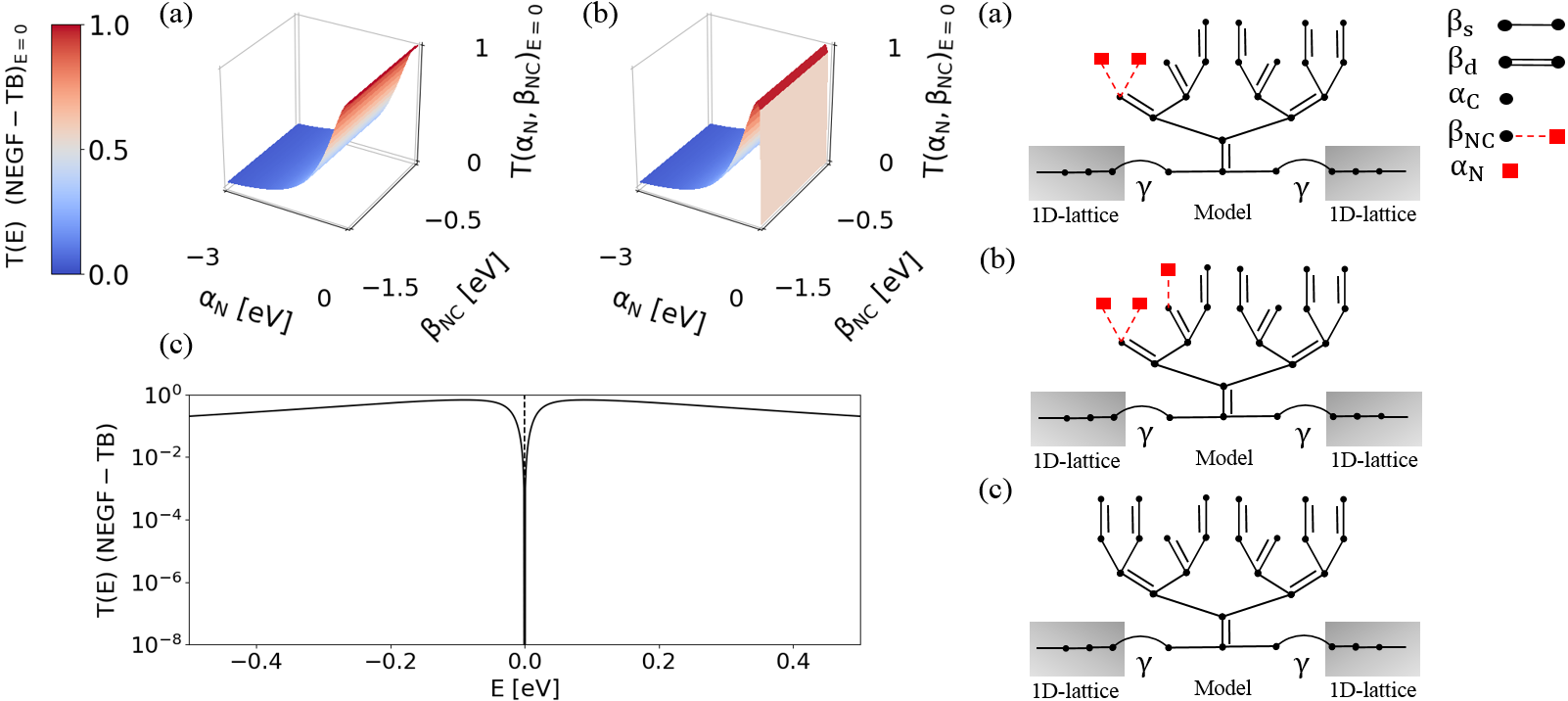} 
\caption{NEGF-TB transport calculations for $3^{\text{rd}}$ generation quantum $\texttt{NAND}$ trees. Calculation Details: Transmission is calculate from the NEGF formalism, $T(E) = \text{Tr}\big\{ \mathbf{\Gamma}_L(E) \mathbf{G}^{ret}(E) \mathbf{\Gamma}_R(E) \mathbf{G}^{adv}(E)  \big\}$, and performed using the TB model for the electronic structure. Parameters for the 1D-lattice: $\alpha_{\text{1D}} = 0$ eV and $\beta_{\text{1D}} = -1.0$ eV. Coupling elements between non-neighbour states are zero. Parameters for model: $\beta_s = -2.4$ eV, $\beta_d=-2.7$ eV, and $\alpha_C := 0$ eV\cite{Purcell}. The coupling element between the model and the 1D-lattice (electrodes) is $\gamma = -1.0$ eV.}
\label{model2}
\end{figure}

\hfill \break 
Fig. \ref{model1}(b), a conjugated molecule, represents a quantum \texttt{NAND} tree, analogous to the classical \texttt{NAND} tree in Fig. \ref{model1}(a) by having the same inputs. Here, the 1D-lattice is replaced with a carbon chain. We require the transmission coefficient at $E=0$ eV for the molecule in Fig. \ref{model1}(b) to be close to one in order to measure the correct final \texttt{NAND} output (1). However, the structure of the \texttt{NAND} tree forces us to add two nitrogen atoms at the leaves in order to avoid unstable carbon radicals and, consequently, two site energies will differ from the rest of the \texttt{NAND} tree. As we saw in Fig. \ref{table:theory}, changing two site energies for leaf structure A will affect the \texttt{NAND} output.
\hfill \break
\hfill \break
We use the non-equilibrium Green's function (NEGF) formalism to calculate the transmission with a 1D-lattice as the electrodes, following from prior work\cite{Reuter2009}\cite{Brisker2008}. The transport calculations are performed using the TB model for the electronic structure, and we label the method as NEGF-TB. Fig. \ref{model2}(a) and (b) show transmission as a function of the coupling element between nitrogen and carbon, $\beta_{NC}$, and the site energy for nitrogen, $\alpha_{N}$, at constant energy $E=0$ eV, and Fig. \ref{model2}(c) shows transmission as a function of incoming energy [eV].

\begin{figure}[H] 
\centering  
\includegraphics[width=0.70\textwidth]{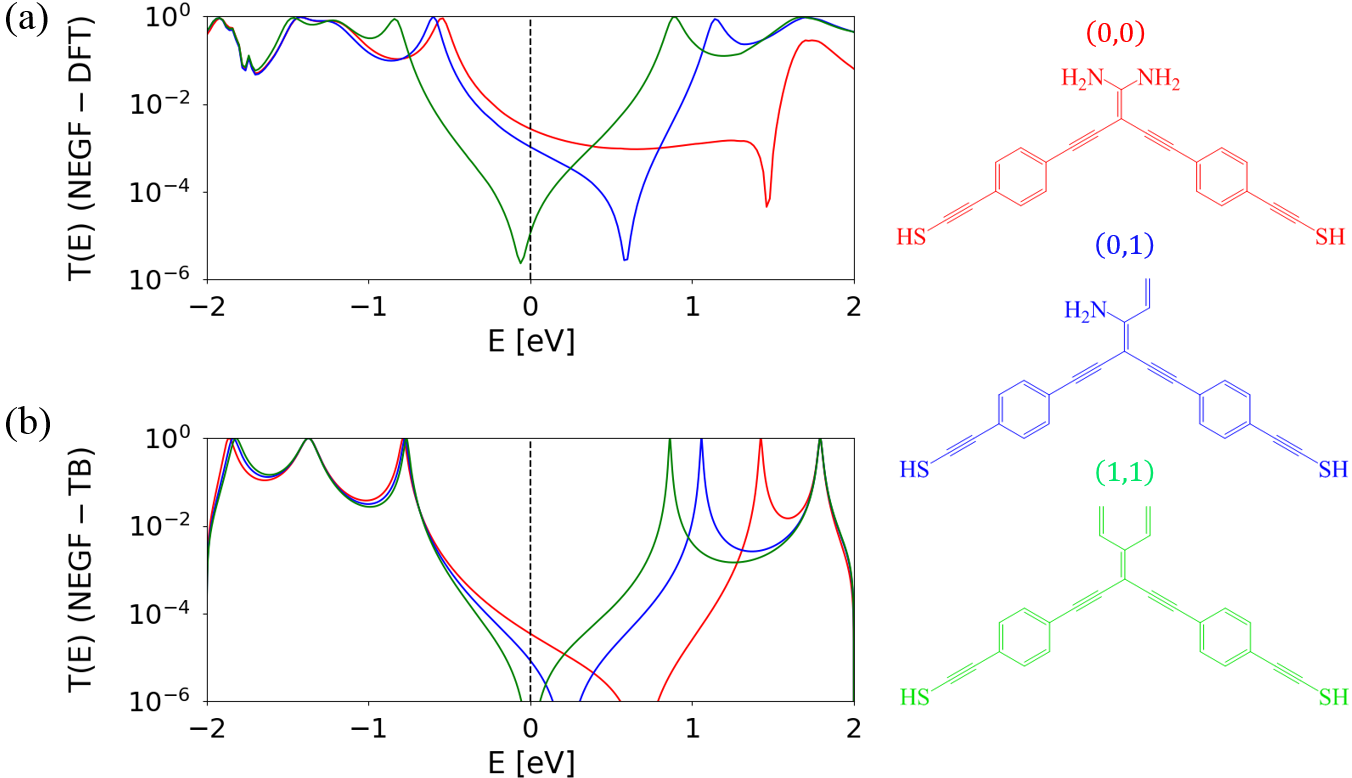} 
\caption{NEGF-DFT and NEGF-TB transport calculations for cross-conjugated molecules that represent $1^{\text{st}}$ generation $\texttt{NAND}$ trees. Calculations Details. (a) Transmission is calculated using the NEGF formalism, $T(E) = \text{Tr}\big\{ \mathbf{\Gamma}_L(E) \mathbf{G}^{ret}(E) \mathbf{\Gamma}_R(E) \mathbf{G}^{adv}(E)  \big\}$, and performed using the DFT Hamiltonian for the electronic structure. To model exchange and correlation effects the Perdew–Burke-Ernzerhof (PBE) functional\cite{Perdew1996} was used, and the wavefunctions were expanded using double-$\zeta$ polarized (DZP) basis sets. We used the code from QuantumWise atomistix toolkit (ATK). (b) Transmission is calculate from the NEGF formalism, and performed using the TB model for the electronic structure. Parameters for the 1D-lattice: $\alpha_{\text{1D}} = 0$ eV and $\beta_{\text{1D}} = -1.0$ eV. Coupling elements between non-neighbour states are zero. Parameters for model: $\beta_{C-C} = -2.4$ eV, $\beta_{C=C}=-2.7$ eV, $\beta_{C \equiv C}=-3.0$ eV,  $\beta_{C \overline{\ldots} C}=-2.55$ eV, $\beta_{NC}=-1.08$ eV, $\beta_{SC}=-2.16$ eV, $\alpha_N = -2.7$ eV, $\alpha_S =-4.05$ eV and $\alpha_C \equiv 0$ eV, as suggested in the literature\cite{Purcell}. Coupling elements between non-neighbour states are zero. The coupling elements between the model and the 1D-lattice (electrodes) are $\gamma = -1.0$ eV. }
\label{fig. DFT}
\end{figure}

\hfill \break
The \texttt{NAND} tree in Fig. \ref{model2}(a) has the final \texttt{NAND} output ($y=0$), which should translate to a high transmission at $E = 0$ eV (see Eq. \ref{Eq.1}). When the site energy for nitrogen equals the site energy for carbon, $\alpha_N = \alpha_C = 0$ eV, the particle is fully transmitted as expected from QST (Fig.  \ref{table:theory}), and we measure the final \texttt{NAND} output ($y=0$). When the site energy for nitrogen differs from carbon (which is the case in reality), $\alpha_ N \neq 0$ eV, the transmitted amplitude decreases, also expected from QST, because we are varying site energies that affect the transmission coefficient. As we approach $\alpha_ N = -3.0$ eV, we see that $T = 0.04$. FIG. \ref{model2}(b) differs in its input by having one more leaf bonded to a quantum state. The change in input, $(Y'=-\infty,Y''=0) \rightarrow (Y'=0,Y''=0)$, results in the final \texttt{NAND} output ($y = - \infty$) which we measure with a low transmission at $E = 0$ eV. When the site energy for nitrogen equals the site energy for carbon, $\alpha_N = \alpha_C = 0$ eV, the particle is fully reflected, $ T\sim 10^{-30}$, as expected from QST, otherwise for $\alpha_ N \neq 0$ eV, the transmitted amplitude starts to increase. At $\alpha_ N = -3.0$ eV, however, the transmission has decreased again to $T =0.04$. The transmission ratio between the \texttt{NAND} trees (a) and (b) is 1, thus we cannot distinguish between the outputs (0) and (1) in this limit. Fig. \ref{model2}(c) also has the final \texttt{NAND} output ($y = - \infty$) but since the \texttt{NAND} tree consists of equal site energies only, destructive quantum interference (QI) is observed at $E=0$ eV, as expected from QST, given very low transmission as required. Here, the transmission ratio between the \texttt{NAND} trees (a) and (c) is large and one can easily distinguish between the outputs (0) and (1).
\hfill \break
\hfill \break
Thus, for \texttt{NAND} trees where variation in site energies affect the final \texttt{NAND} output, as in Fig. \ref{model2}(a) and (b), the magnitude of the site energy variation will determine whether we can distinguish the outputs (0) and (1). While this results from both a reduction in the high-transmission case and an increase in the low-transmission case, the dominant contribution comes from the many orders of magnitude increase in the low transmission. The reason is that the destructive QI effects at $E=0$ very rapidly looses its ability to suppress transmission with variation in site energies, as we saw in Fig. \ref{model2}(b) with $T_{\alpha_N = 0} \sim 10^{-30} \rightarrow T_{\alpha_N = -3} \sim 10^{-3}$ just by a few eV. Destructive QI effects in acyclic cross-conjugated systems, like the structures employed here, are sensitive to variation in site energies\cite{Solomon2008}, thus, in this realisation, performance is significantly improved in \texttt{NAND} trees with equal site energies, as in Fig. \ref{model2}(a) and (c).
\hfill \break
\hfill \break
TB model calculations are, however, only a very approximate description of the molecular electronic structure. We can improve this description by using density function theory (DFT), allowing us to include both $\sigma$ and $\pi$ electrons as well as a realistic non-planar geometry for the molecule. Fig. \ref{fig. DFT} shows the comparison between the NEGF-DFT and NEGF-TB for $1^{\text{st}}$ generation $\texttt{NAND}$ molecules with linker groups to simulate the binding of these molecules to metal electrodes. The labels (0,0), (0,1), and (1,1) indicate the inputs, hence the red and blue molecules should have the final output (1) and the green molecule the final output (0), which is achieved with a high and a low transmission, respectively. 

\hfill \break
The molecule (1,1) consists of carbons only, thus all site energies are the same, $\alpha_C \equiv 0$ eV, and destructive QI is observed at $E=0$ eV given very low transmission as required. In the NEGF-DFT transport calculations, the destructive QI features are also preserved during stretching and compression of the Au-S bond (see supporting information Figure S4). On the other hand, the inclusion of the nitrogen atoms clearly shifts the resonances away from the Fermi energy compared with all carbon systems, as expected from QST, so the high transmission required for the molecules (0,0) and (0,1) is not so successfully achieved. The NEGF-TB ratios are large, $T_{(0,0)}^{TB}/T_{(1,1)}^{TB} \sim T_{(0,1)}^{TB}/T_{(1,1)}^{TB} \sim 10^{25}$, see Table \ref{tabel2}, thus the bit values $\{0,1\}$ are preserved within the TB model. The NEGF-DFT ratios, $T_{(0,0)}^{DFT}/T_{(1,1)}^{DFT}$ and $T_{(0,1)}^{DFT}/T_{(1,1)}^{DFT}$, are 250 and 100 respectively, thus the bit values are preserved to some extent but reduced to a on/off transmission ratio.

\begin{figure}[H] 
\centering  
\includegraphics[width=0.88\textwidth]{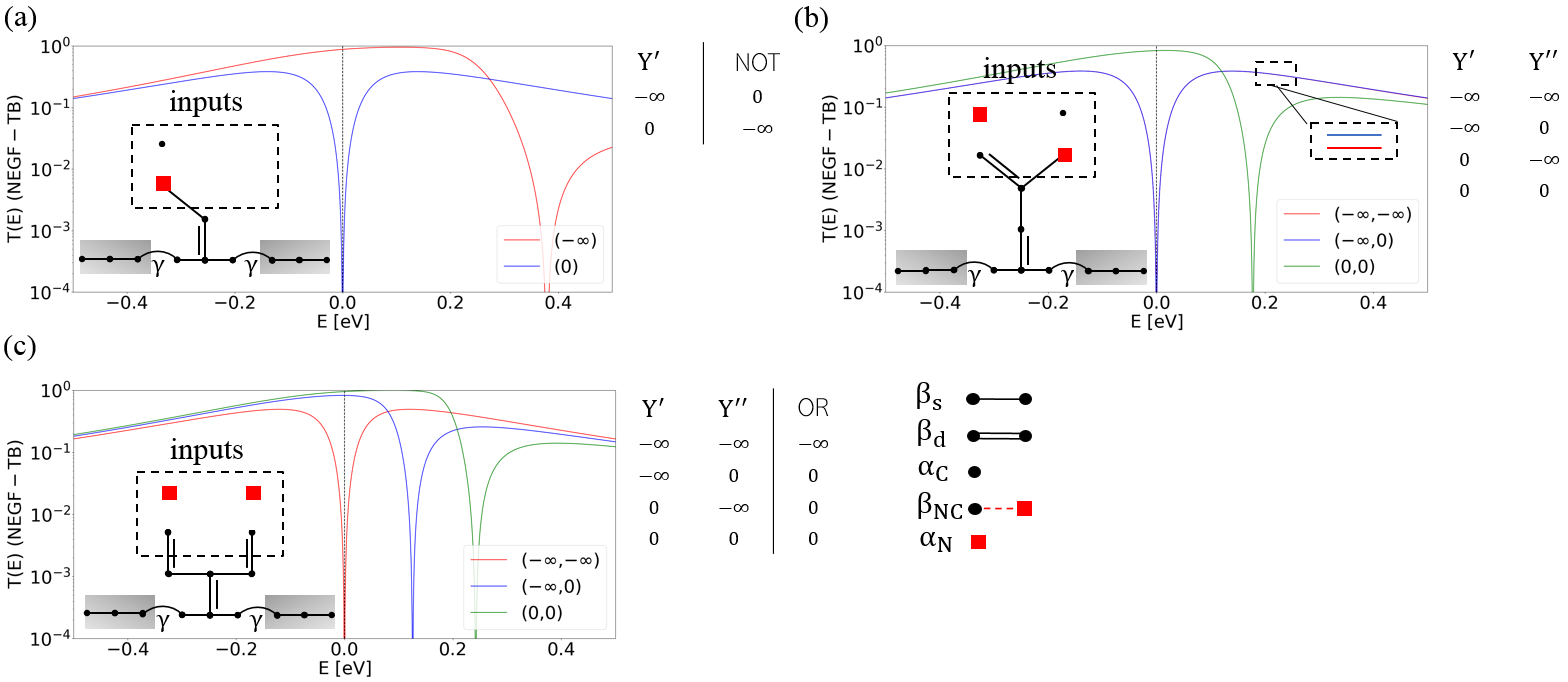} 
\caption{NEGF-TB transport calculations for $1^{\text{st}}$ generation quantum \texttt{NOT}, \texttt{AND}, and \texttt{OR} gates. Calculation Details: Transmission is calculated from the NEGF formalism, $T(E) = \text{Tr}\big\{ \mathbf{\Gamma}_L(E) \mathbf{G}^{ret}(E) \mathbf{\Gamma}_R(E) \mathbf{G}^{adv}(E)  \big\}$, and performed using the TB model for the electronic structure. Parameters for 1D-lattice: $\alpha_{\text{1D}} = 0$ eV and $\beta_{\text{1D}} = -1.0$ eV. Coupling elements between non-neighbour states are zero. Parameters for model: $\beta_{C-C} = -2.4$ eV, $\beta_{C=C}=-2.7$ eV, $\beta_{NC}=-1.08$ eV, $\alpha_N = -2.7$ eV, and $\alpha_C \equiv 0$ eV\cite{Purcell}. Coupling elements between non-neighbour states are zero. The coupling elements between the model and the 1D-lattice (electrodes) are $\gamma = -0.5$ eV.}
\label{fig. OtherGates}
\end{figure}

\begin{center}
\captionof{table}{Transmission coefficients at $E=0$ eV for molecules in Fig. \ref{fig. DFT}. Column 1: The inputs for the molecules. Column 2: The truth table for a classical \texttt{NAND} gate (not included (1,0)). Column 3: Transmission at Fermi-level calculated with NEGF-DFT. Column 4: Transmission at Fermi-level calculated with NEGF-TB.}
\begin{tabular}{l*{10}{c}r}
\hline
\hline  \\[-1.5ex]
Molecule && A & B &$\overline{\text{A$\cdot$B}}$ && $T^{DFT}(E)_{E-E_f=0}$ && $T^{TB}(E)_{E-E_f=0}$ \\[1.5ex]
\hline \\
(0,0)   && 0 & 0 & 1 && 2.7 $\cdot$ $10^{-3}$  && 4.1 $\cdot$ $10^{-3}$ \\[0.5cm]
(0,1)   && 0 & 1 & 1 && 1.1 $\cdot$ $10^{-3}$ && 1.0 $\cdot$ $10^{-3}$   \\[0.5cm]
(1,1)  && 1 & 1 & 0 && 1.1 $\cdot$ $10^{-5}$ && 9.8 $\cdot$ $10^{-28}$   \\[0.2cm] \hline 
\hline
\vspace{0.1cm}
\end{tabular}
\label{tabel2}
\end{center}
The same approach can be used to build other classical gates for quantum computing. Fig. \ref{fig. OtherGates} shows the classical gates, \texttt{NOT}, \texttt{AND}, and \texttt{OR}, their quantum structures, and NEGF-TB transport calculations to show their final outputs. The labels refer to the input for the logic gate, for example the label ($-\infty$) for the \texttt{NOT} gate has input($Y'=-\infty$) and output($Y=0$) which we measure with a high transmission at $E = 0$. The inputs are given at the leaves and we control the inputs by connection/disconnection states, as we did in the quantum \texttt{NAND} tree.

\section{Conclusion and Perspectives}
In conclusion, by analyzing the quantum \texttt{NAND} tree, we showed the restrictions on the TB parameters required to maintain \texttt{NAND} gate function. An important result was the ability to vary coupling elements without affecting the \texttt{NAND} gate function, which indeed is a necessary condition when moving from the model to a molecule. The site energies could also be varied to some extent but there are significant restrictions on which site energies can vary while maintaining the \texttt{NAND} gate function. Further, we mapped the quantum \texttt{NAND} tree onto a conjugated molecule. Due to the structure of the \texttt{NAND} tree, we replaced some carbon atoms with nitrogen atoms, and, consequently, the destructive QI at Fermi-level very rapidly looses its ability to suppress transmission. Consequently, in some cases, the \texttt{NAND} outputs (0) and (1) were not successfully achieved. Beyond the issues associated with finding a suitable molecule, there are further physical processes that could well limit the function of a molecular \texttt{NAND} gate utilizing quantum interference. Namely:
\hfill \break
\hfill \break
1) \textit{Dephasing}:
Prior work studying charge transfer\cite{Goldsmith2006}\cite{Goldsmith2007} and transport\cite{Andrews2008} has shown that destructive interference effects in molecules will indeed be suppressed (i.e. transmission increases) when dephasing is included in the calculations, this will effect the off state for the \texttt{NAND} gates. In all these cases, pure dephasing was introduced with a parameter, not from an \textit{ab initio} calculation, and a realistic value for this parameter at a particular temperature is unknown. 
\hfill \break
2) \textit{Disorder/Geometric fluctuations}:
Even if we assume the electron transfer event is sufficiently fast (or instantaneous) such that pure dephasing is not a concern, we need to consider the effect of geometric fluctuations, both within the molecule and of the molecule in the junction. These kinds of fluctuations have been modeled by taking an ensemble of geometries from molecular dynamics simulations and calculating the transmission\cite{Andrews2008}. In that case, it was seen that these sorts of fluctuations will shift the peaks and dips in transmission in energy (effectively noise in the energy scale of the transmission plots) but sharp interference dips remain.
\hfill \break
3) \textit{Inelastic transport}:
The final aspect to consider is whether transport processes beyond the elastic transport we have considered here could be significant, in particular, what is the magnitude of inelastic transport? Theoretical studies have shown that while there may certainly be significant inelastic contributions to the current, the pronounced suppression as a result of destructive interference is not completely removed\cite{Lykkebo2014}\cite{Markussen2014}.
\hfill \break
\hfill \break
All the studies listed above were theoretical, and none considered all aspects at once. Consequently, the question still remains as to whether the sum total of these processes might present a problem, even if each one alone does not. At this stage this has not been quantified, but the extremely small size of the molecular components and the fact that significant interference effects have been observed experimentally at room temperature\cite{Garner2018}\cite{Guedon2012}, suggests that destructive interference effects persist despite these physical processes.
\hfill \break 
\hfill \break 
The choice of an acyclic cross-conjugated system for the molecular realization of the quantum \texttt{NAND} tree yields a molecule with a H\"uckel Hamiltonian that was as similar as possible to the tight binding model from the original work. The problem however, is that not all of the tight-binding systems correspond to stable molecular systems without modification. The general principle of the quantum \texttt{NAND} tree, that QI interference effects can be switched on and off with substituents and thereby give \texttt{NAND} function, can of course be explored with many other systems known to result in destructive QI. Many systems have been explored for this purpose, for example, meta-substituted benzene rings, five-membered rings, and quinoidal structures\cite{Garner2016}\cite{Borges2017}. It remains an open question as to whether any of these systems would offer improved performance as a \texttt{NAND} gate, in particular offering a larger transmission ratio for the final \texttt{NAND} outputs. 
\hfill \break 
\hfill \break 
Another challenge that remains is how best to change the logical inputs for the quantum \texttt{NAND} tree. Here, this was achieved by connection/disconnection of substituents, which is not suitable for experiments. Soe et al.\cite{Soe2011} have conducted scanning tunneling microscopy (STM) experiments  realizing Boolean truth tables through molecules, including changing the logical inputs. In that case, their synthesized molecule is physisorbed on a Au(111) surface (the $\pi$-system is preserved), and current tunnels from the STM tip, through the molecule, to the surface. As in the quantum \texttt{NAND} tree, the logic gate function is encoded in the transmission, and they control the inputs by contacting or detaching Au atoms at the top of the molecule.
\hfill \break 
\hfill \break 
While a practical realization of the quantum \texttt{NAND} tree in a molecular system remains elusive, it is clear that there is a viable strategy for future investigation. The molecular nature of quantum interference effects in molecules, with their observation in many molecular systems at room temperature, suggest that molecules should be ideal for this purpose $-$ if the right molecular structures can be found. Chemical space is certainty vast but until recently chemists have not focused on developing chemical insight for these sort of applications. As research into quantum interference effects in molecules progresses, so too may we see the emergence of molecular candidates for quantum realization of classical logic gates, or even quantum gates.

\section*{Acknowledgements}
We thank Magnus F. B{\o}e, Anders Jensen, Marc H. Garner, Idunn Prestholm, Thorsten Hansen and Christian Joachim for helpful discussions. This research was supported by the Danish Chemical Society, Oticon Fund, Julie Damms Fund, Copenhagen Education Fund, Knud H{\o}jgaards Fund, Henry and Mary Skovs Fund, Viet - Jacobsens Fund, Villum Foundation, Carlsberg Foundation, the Danish Council for Independent Research $\mid$ Natural Sciences, and the Vannevar Bush Faculty Fellowship program sponsored by the Basic Research Office of the Assistant Secretary of Defense for Research and Engineering and funded by the Office of Naval Research through grant N00014-16-1-2008.

\clearpage
\section*{References}
\bibliographystyle{ieeetr}
\bibliography{library}

\end{document}